\documentstyle[amssymb,aps,preprint,prc,psfig]{revtex}

\begin{document}
\author{A.Chernomoretz, M. Ison, S.Ortiz and C.O.Dorso}
\date{\today}
\title{Non equilibrium effects in fragmentation}
\maketitle

\begin{abstract}
We study, using molecular dynamics techniques, how boundary conditions
affect the process of fragmentation of finite, highly excited, Lennard-Jones
systems. We analyze the behavior of the caloric curves (CC), the associated
thermal response functions (TRF) and cluster mass distributions for
constrained and unconstrained hot drops. It is shown that the resulting CC's
for the constrained case differ from the one in the unconstrained case,
mainly in the presence of a ``vapor branch''. This branch is absent in the
free expanding case even at high energies . This effect is traced to the
role played by the collective expansion motion. On the other hand, we found
that the recently proposed characteristic features of a first order phase
transition taking place in a finite isolated system, i.e. abnormally large
kinetic energy fluctuations and a negative branch in the TRF, are present
for the constrained (dilute) as well the unconstrained case. The microscopic
origin of this behavior is also analyzed.
\end{abstract}


\address{Departamento de F\'{\i}sica, Facultad de Ciencias Exactas y\\
Naturales Universidad de Buenos Aires\\
Pabellon I, Ciudad Universitaria, Nu\~nez\\
1428 Buenos Aires, Argentina}

\newpage

\section{\protect\smallskip Introduction}

The possibility of getting information about the thermodynamics of nuclear
matter from the analysis of intermediate energies heavy ion collisions has
triggered a lot of interest in this field. Starting from the pioneering work
of the Purdue group, when for the first time a power law was used to fit the
mass spectra resulting from highly excited nuclear systems and thus
suggesting that a critical phenomena was taking place, it has open the study
of a rather new branch of thermodynamics i.e the study of phase transitions
(liquid-vapor) in finite systems\cite{nuovo}.

One of the most challenging features that nuclear multi-fragmentation
phenomena presents, is that signals of phase transitions of different orders
can be extracted from experimental data. On one hand, the analogy between
the nuclear force and a van der Waals interaction suggests that the nuclear
equation of state should reproduce the main features that characterize a
liquid-gas phase transition. Different indications of this kind of
transition have been reported \cite{pocho95,gross,gulminelliDurand,dagostino}%
. On the other, signatures of a second order phase transition or critical
behavior have been obtained assuming Fisher-like scaling relations for
fragment distributions~\cite{purdue,jbelliot,balen,campi,bauer}. Moreover,
in recent works \cite{gulminelli,carmona} is suggested that the observed
critical behavior is compatible with a first order phase transition, and it
is due exclusively to finite size effects.

An important point to be kept in mind is that the above mentioned approaches
to the fragmentation problem are based on descriptions where not only the
concept of thermodynamical equilibrium, but also the macroscopic constraints
imposed to the system play a relevant role. For example, several statistical
descriptions of the nuclear multi-fragmentation process, e.g. the
statistical multifragmentation model (SMM)~\cite{SMM} and the microcanonical
Metropolis Monte Carlo model (MMMC)~\cite{gross}, employ the concept of a 
{\it freeze-out volume} inside of which the existence of a thermodynamical
equilibrated ensemble of fragments is assumed. These statistical models have
been widespreadly used by the nuclear community in a successful manner to
described some experimental observations. However, from the experience
gained in numerical simulations, such a concept as {\it freeze-out volume},
even as an approximation, does not seem to be completely correct. It is then
important to study which kind of differences arise as a consequence of not
assumming a finite volume scenario, as it is not {\it a priori} evident
that the evolution of a fragmenting system confined in a finite volume would
produced the same macroscopic observable when compared with a non confined
one, in which an expansive motion is present as an extra collective degree
of freedom.

In previous works~\cite{ale_time,ale_temp} we have studied the fragmentation
of a simple classical system where the dynamics is governed by a Hamiltonian
with a two body interaction Lennard-Jones term. A microscopic description,
employing molecular dynamics techniques, was used in order to adequately
handle the possible presence of a non equilibrium behavior. It was shown
that a fragmentation time can be defined, after which a certain degree of
local equilibrium is achieved in the system. This fact allowed us to
calculate a caloric curve for our expanding-fragmenting system, which is
characterized by the absence of a vapor branch.

The aim of the present communication is to study how the restriction of a
finite volume, and then the imposition of equilibration, affects some of the
results obtained in the unconstrained case. We will show that one of the
main effects is seen in the behavior of the caloric curve (CC). For the
constrained system it clearly shows a vapor-branch, which is absent in the
free expanding case. This behavior is a direct consequence of the presence
of a confining volume, which destroys the velocity correlations that, in the
case of the free expanding system, build up an expansive flux that acts as a
heat sink. Nevertheless for both cases (at rather low densities for the
constrained system) a local maximum and a loop in the CC can be seen. This
feature can be associated with a negative branch in the corresponding
thermal response function (TRF), that might signal a phase transition of
first order. A gradual smoothing of the mentioned indicators is observed for
the confined system at higher densities. Differences in fragment mass
distributions are also reported and related to the microscopic origin of
such behavior.

This paper is organized as follows. In Section~\ref{NumSim} we will describe
the model used in our simulations. A brief review of the results already
obtained for the unconstrained expanding system is included. Section~\ref
{secCC} is devoted to the study of the caloric curves. In Section~\ref
{secVfo} we study possible choices for the freeze out volume for the
constrained case. In Section~\ref{secTRF} we calculate the thermal response
functions and the kinetic energy fluctuations in order to signal the
presence of possible phase transitions. A microscopic correlations study is
performed in Section~\ref{secSpectra} analyzing the results of two different
clusterization algorithms. Finally, in Section~\ref{secConclu}, conclusions
are drawn.

\section{Numerical simulations}

\label{NumSim}

\smallskip The system under study is composed by excited drops made up of
particles interacting via a 6-12 Lennard Jones potential, which reads:

\begin{equation}
V(r)=\left\{ 
\begin{array}{ll}
4\epsilon \left[ \left( \frac{\sigma }{r}\right) ^{12}-\left( \frac{\sigma }{
r}\right) ^{6}-\left( \frac{\sigma }{r_{c}}\right) ^{12}+\left( \frac{\sigma 
}{r_{c}}\right) ^{6}\right] r \le r_{c} &  \\ 
0 r$>$r_{c} & 
\end{array}
\right.
\end{equation}

\smallskip We took the cut-off radius as $r_{c}=3\sigma $. Energies and
distances are measured in units of the potential well ($\epsilon $) and the
distance at which the potential changes sign ($\sigma $), respectively while
the unit of time used is: $t_{0}=\sqrt{\sigma ^{2}m/48\epsilon }$. We
integrated the set of classical equations of motion using the well known
Verlet algorithm \cite{Verlet}, taking $t_{int}=0.002t_{0}$ as the
integration time step. Initial conditions were constructed using the already
presented method of cutting spherical drops composed of 147 particles out of
equilibrated and periodic 512 particles per cell L.J. system \cite{ale_temp}.

A broad energy range was considered such that the asymptotic mass spectra of
the fragmented drops for the unconstrained system changes from a ''U
shaped'' pattern to an exponentially decaying one. Somewhere in between this
two extremes a power law like spectra can be obtained.

\subsection{Analysis of the unconstrained expanding case}

\label{SFES}

For the sake of completeness we summarize in this section the main results
obtained in previous work for the non-constrained fragmenting system (see 
\cite{ale_temp} for details).

Due to the fact that we are dealing with a non-stationary process the
determination of the time at which fragments are formed becomes one of the
key ingredients in the analysis. To accomplish this task one can use a
simple and intuitive cluster definition, that is based on correlations in
configuration space: a particle $i$ belongs to a cluster $C$ if there is
another particle $j$ that belongs to $C$ and $| {\bf r_i}-{\bf r_j}| \leq
r_{cl}$, where $r_{cl}$ is a parameter called clusterization radius (In this
work we used $r_{cl}=r_{cut}=3\sigma$). The recognition algorithm introduced
by this definition is known as minimum spanning tree (MST) fragment
recognition method.

In \cite{ale_mstecra} it was shown that a different recognition method , the
Early Cluster Formation Model and its companion practical realization the so
called Early Cluster Recognition Algorithm (ECRA), outperforms the MST
clusterization algorithm in the sense that it finds the asymptotic
fragmentation pattern in phase space at early stages in the evolution, when
fragments are still not observables in configuration space. At that time the
system still looks like a rather compact piece of excited matter. Instead of
being defined through a proximity criteria, the ECRA fragments are
associated with the set of clusters $\{C_{i}\}$ for which the sum of the
fragment internal energies attains its minimum value: 
\begin{eqnarray}
{\{C_{i}\}} &&{=}{\hbox{min}}{{\scriptstyle \{C_{i}\}}}{\textstyle {%
[E_{\{C_{i}\}}=\sum_{i}E_{int}^{C_{i}}]}}  \nonumber \\
E_{int}^{C_{i}} &=&\sum_{i}[\sum_{j\in C_{i}}K_{j}^{c.m.}+\sum_{{j,k\in C_{i}%
}{j\le k}}V_{j,k}]  \label{eq:eECRA}
\end{eqnarray}

where the first sum in (~\ref{eq:eECRA}) is over the clusters of the
partition, $K_{j}^{c.m.}$ is the kinetic energy of particle $j$ measured in
the center of mass frame of the cluster which contains particle $j$, and $%
V_{ij}$ stands for the inter-particle potential. We dub the partition found
by ECRA as the most bound density fluctuation in phase space (MBDF) and we
define as time of fragment formation ($\tau _{ff}$) the time at which the
MBDF attain microscopic stability (see appendix for details). In this way $%
\tau _{ff}$ is related to the time at which the systems switches from a
regime dominated by fragmentation to one in which the dominant decay mode is
evaporation of light aggregates (mostly single particles) by the excited
fragments. In this way the ECFM-ECRA outperforms the MST not only in terms
of the time at which fragments are detected but also in it's capability of
unveiling the nature of fragmentation process. It shows that fragments are
formed in phase space as a consequence of correlations in both q-p space.

Once the time of fragment formation is determined it is possible to
calculate several properties of the system at fragmentation time. For this
purpose the expanding system is decomposed in concentric shells and the mean
radial velocity is calculated. It can be seen that the expansion is almost
linear with the distance to the center of mass of the system, and that a
local temperature can be defined as the fluctuations of the velocity around
the local expansive collective motion. Moreover, the isotropical character
of those fluctuations supports the idea that local equilibrium is achieved 
\cite{ale_temp}.

It is worth noticing that, on one hand at $\tau _{ff}$ most of the system is
still interacting, and on the other the local temperature of the inner
shells attain a rather constant value that can be consistently considered as
the temperature of the system at fragmentation time. This can be seen in
figure~\ref{figtloc} where we show, for different energies, the dependence
of the local temperature, $T_{loc}(r)$, and the local density, $\rho (r)$,
as a function of the distance to the center of mass at fragmentation time.

\subsection{Analysis of the Constrained System}

In order to study the consequences of imposing a finite volume constraint to
our system we used a spherical confining `wall'. The considered external
potential behaves like $V_{wall} \sim (r-r_{wall})^{-12}$ with a cut off
distance $r_{cut}=1\sigma $ , where it smoothly became zero along with its
first derivative. A rather broad range of values for $r_{wall}$ were used.

Inside this potential, a highly excited drop was initialized in the way
already described above and the corresponding equations of motion were
integrated. Once the transient behavior was over we performed a
microcanonical sampling of particle configurations every $5t_{0}$ up to a
final time of $140000t_{0}$.

In this case the standard prescription for temperature calculation in the
microcanonical ensemble was used, i.e. the kinetic energy ($K$) was related
to the temperature of our $N$-particles system using: 
\[
T={\frac{2}{{3(N-1)}}}K 
\]

\section{Caloric curves}

\label{secCC} One of the thermodynamics measurements that remains useful in
the small system limit (hundred of particles) is the caloric curve, i.e. the
functional relationship of the system temperature with its excitation energy.

\smallskip The resulting caloric curve (CC) for the unconstrained system is
displayed in Fig.~\ref{figCC}~a). In this figure circles denote the
temperature of the system, already defined in Section~\ref{SFES}, as a
function of the energy. On the same figure a curve denoted by squares is
also present, for this one the total kinetic energy, i.e. including both the
fluctuations around the collective motion and the collective motion itself,
is used instead. Of course this is not a temperature , just a fraction of
the total kinetic energy, but it is then obvious that collective motion
begins to be noticeable around a value of $E=0$ , and it becomes dominant at
around $E=2.\epsilon $. Two main features are to be noticed, in first place
the CC develops a maximum, and second the CC has no ''vapor branch'' but
develops a rather constant behavior. These features are in contrast with the
standard expectations inherited from the thermodynamics of infinite systems.

In Fig.\ref{figCCV3D}) we show the behavior of $T(E,\rho )$ for the
constrained system (see also Fig.\ref{figCCV}a), where for the sake of
clarity, we include just three different values of $\rho $).

For all density values a vapor branch is clearly observed. Moreover,
comparing Fig.\ref{figCC}a) with Fig.\ref{figCCV3D}) and~Fig.\ref{figCCV}a),
it is clear that the vapor branch is due to the presence of the confining
volume which inhibits the formation of radial collective motion that behaves
like a heat sink in the non constrained case. Inside the volume the emitted
aggregates can do nothing but interact among themselves, and with the
constraining wall, until thermal equilibrium is attained.\newline

The effect of the boundaries on the CC is not important as long as the total
energy in the system is small enough, and the expansive collective motion
(in the unconstrained case) can be neglected. But this is not the case for
highly excited finite systems, for which the presence of the constraining
wall prohibits the formation of the expansive radial motion. For that range
of energies the local equilibrium features that can be found in the
fragmentation of the unconstrained system are replaced by global equilibrium
ones when the dynamics is confined to finite volumes. This means that one
has to be at least extremely careful if one intends to analyze expanding
systems using confining volumes or any other model that rely on a global
equilibrium hypothesis.

Another interesting feature to be notice in Fig.\ref{figCCV3D}) and~Fig.\ref
{figCCV}a) is the presence of a loop at the beginning of the gas rise in the
CC for diluted enough systems. This back-bending behavior is gradually
smeared out as the system density is increased. In Section~\ref{secTRF} we
will relate this to the behavior of the thermal response function of the
system and in Section~\ref{secSpectra} we will find that this can also be
understood with the aid of the mass spectrum resulting from the MST and ECRA
analysis.

\section{Freeze out volume approximation}

\label{secVfo}

As was already shown, due to the intrinsic non-equilibrium character of the
fragmentation process in the non-confined case, an assumption of local
equilibrium (`mounted' over an expansive radial flux) instead of a global
one has been found to be more appropriate. Accordingly, distributions of
density and temperature, and not unique values, are necessary to described
the system at fragmentation time in a rigorous way(see Fig.\ref{figtloc}).
It is then not expected to find an exact mapping in the ($T,\rho $) space
between the constrained and non-constrained dynamics.

Nevertheless if one insists on establishing such a comparison, two different
criteria can be adopted in order to choose the appropriate volume for the
constrained case (which we will call {\it freeze-out volume}, $V_{fo}$). On
one hand it can be seen from Fig.\ref{figtloc}) that at $\tau _{ff}$ the
local density value for the inner cells (where most of the mass is present
at the time of fragment formation) remains rather constant ($\rho \sim
0.08\sigma ^{-3}$) for the broad energy range presented in the figure.
Accordingly, one can choose $V_{fo}$ in order to attain such density value
in the constrained case. This election, that could be considered as a
`density-guided' choice, corresponds to $r_{wall}\sim 8\sigma $ and a {\it %
freeze out} density value of $\rho _{fo}^{+}\sim \rho _{0}/10$ .

For the other hand a different approach can also be adopted looking at Fig.%
\ref{figCC}a) and~Fig.\ref{figCCV}a). It can be verified that for energy
values where the onset of the fragmentation process occurs ($-2\epsilon \le
E\le 0\epsilon $) the thermal energy in the expanding system is clearly
lower than for the $r_{wall}=8\sigma $ case in the constrained case. In
fact, similar temperatures can be found in the constrained case only for
much diluted situations, i.e. for $r_{wall}=12\sigma $. In this sense, from
a `temperature point of view', a different {\it freeze out} density $\rho
_{fo}^{-}\sim 0.02\sigma ^{-3}\sim \rho _{0}/40$ can be established.

As we mentioned above this ambiguity in the definition of a {\it freeze out}
volume (or density) is a consequence of trying to reproduce the behavior of
an expanding fragmenting system using a global equilibrium scenario.

\section{Thermal response function}

\label{secTRF}

In recent works~\cite{gross,dagostino,chomaz} a lot of attention has been
paid to the role played by the behavior of the specific heat (or more
generally speaking: the thermal response function, $TRF$) as a signal of the
occurrence of a phase transition in finite systems. Moreover, it has been
shown (see~\cite{gross}) that the presence of a negative branch in the $TRF$
can be related to a first order phase transition taking place in an isolated
finite system. This kind of analysis can straightforwardly be performed over
the systems under the current study, using the already calculated CC and
taking into account that the respective TRF's can be calculated as:

\begin{equation}  \label{trf_cc}
TRF= (\frac{\partial T}{\partial E})^{-1}
\end{equation}

In Fig.\ref{figCC}b) we show the TRF for the unconstrained system. Two poles
can be seen. The first one ($E\sim -0.5\epsilon $) signals the entrance of
the system into the multifragmentation regime, while the second one is
related to the leveling off in the corresponding CC, and can be related to
the increasing limit imposed by the strong flux to the 'thermalization' of
the total available energy.

Fig.\ref{figCCV}b) shows the corresponding TRF for the constrained case. We
include the curves for the two limiting cases discussed in Section~\ref
{secVfo} as possible freeze out choices, i.e. $\rho _{fo}^{+}$ and $\rho
_{fo}^{-}$, and an intermediate $\rho ^{0}\sim \rho _{0}/20$ value ($\rho
_{fo}^{-}<\rho ^{0}<\rho _{fo}^{+}$). We notice that two poles and a
negative branch can be observed for the $\rho _{fo}^{-}$ TRF curve. In this
case we can relate the first pole with the onset of the transition, and the
second one with the entrance in the gas phase, the energy distance between
them being related to the associated latent heat. For $\rho =\rho ^{0}$ it
can be seen that the curve exhibits a qualitatively similar behavior, the
distance between the two poles is just reduced. Nevertheless, for higher
density cases, such as $\rho _{fo}^{+}$, the two poles merge into a single
`singularity' limited by finite-size effects. In that case, the TRF remains
positive for all energies, showing a peak as a signature of the transition,
as a consequence of the fact that the corresponding caloric curve does not
display a loop but instead a simple change of slope (see next section for an
analysis of this change of behavior).\newline

Before leaving this section we would like to discuss, using an approach
introduced in~\cite{chomaz}, the behavior of the fluctuations in the kinetic
energy as an indicator of the occurrence of a phase transition in our finite
system. In~\cite{chomaz} it is shown that for an isolated and equilibrated
system in which the total energy can be decomposed as: $E=E_1+E_2$ the heat
capacity can be calculated as:

\begin{equation}  \label{eqC}
C\approx {\frac{C^2_1 }{C_1 - \sigma^2_1/T^2}}
\end{equation}

where $C_1$ is the heat capacity associated with the subsystem-$1$, $%
\sigma_1 $ is the fluctuation of the partial energy $E_1$, and $T$ the
temperature of the system. In this context, the presence of poles and
negative values in TRF's can be associated to abnormally large fluctuations
of the partial energy stored in subsystem-$1$, i.e.~$\sigma^2_1\ge C_1 T^2$,
during the phase transition.

In Fig.\ref{figFlucV}) we show, for the constrained case, the relative
fluctuation, $A(E)$, for the kinetic energy defined as: 
\begin{equation}
A(E)=N{\frac{\sigma _{K}^{2}|_{E}}{<K>|_{E}^{2}}}  \label{ae}
\end{equation}
where $N$ is the number of particles in the system, and $<>$ stands for an
average over the microcanonical sampling of configurations. We show the
direct calculations using equation~\ref{ae} for both, $\rho _{fo}^{+}$ and $%
\rho _{fo}^{-}$ , and we also include the estimation of $A(E)$ derived for
the respective CC's in Fig.\ref{figFlucV}). A rather good agreement can be
seen between both ways of calculating this magnitude. In the same graph a
reference level marks the canonical value for $A(E)$. As expected, for the
appropriate range of energies, `unusually' large relative fluctuations for
the $\rho _{fo}^{-}$ case can be observed, while for the $\rho _{fo}^{+}$
case a local maximum, that does not exceed the canonical value can be seen.
We postpone an analysis of this behavior until the next section, where we
will study the properties of the system clusters.

We now try to extend the above performed analysis for the unconstrained
system. In order to do that, we considered the central region (or core)
defined by $r<=6\sigma $, and we calculated the total energy, $E_{core}$,
and the number of particles $N_{core}$ that remains inside the core at
fragmentation time. Binning the whole set of events according to these
variables we can sample unconstrained fragmenting events in a {\it %
pseudo-microcanonical} way in what concerns core-quantities at $\tau _{ff}$.
In Fig.\ref{figFluc}) we show the behavior of the relative fluctuation of
the core kinetic energy, $A(E_{core})$, calculated using eq.~\ref{ae} with $%
K=K_{core}$ and $N=N_{core}=60\pm 5$. In the same figure we include the
canonical reference level. Even with a severe reduction of statistics
imposed by the binning, a region of large fluctuations in $K_{core}$, for $%
N_{core}$ and $E_{core}$ fixed, can be easily recognized. This finding
suggest that the 'large kinetic energy fluctuation signal', counterpart of
the behavior observed for the corresponding CC and TRF, can still be found
for the expanding case.

\section{Cluster Distribution}

\label{secSpectra}

In previous sections we have analyzed our system studying thermodynamical
features like the behavior of the CC and the TRF. In this part of the work
we will focus our attention to a different aspect of the process, more
directly related to microscopic correlations, that can be studied using the
clusterization algorithms already introduced in Section~\ref{SFES}.

Let us begin with the simple MST analysis. As this cluster recognition
method is exclusively based on spatial correlations one expects for high
density situations a big MST-cluster to be present. On the other hand, for
lower densities a deviation from the U-shaped behavior in the mass spectra
can be expected. As was already stressed, MST underestimates the number of
clusters because it does not take into account the relative velocities of
the cluster constituent particles. For unconstrained systems this effect was
apparent when we considered the time of fragment formation, at this time the
system was already fragmented in phase space, but in configuration space, a
big cluster was still present. Due to the fact that the system was free to
expand the asymptotic stage was such that MST converged to ECRA. But in the
constrained case such an expansion is prohibited and then MST will not
converge to ECRA.

Nevertheless, an interesting point to be notice is that the MST algorithm
can still provide useful information about the limit imposed by the
constraining finite volume to the formation of well defined fragments in
configurational space because it reflects the size of the interacting
subsystems present at a given time.

To further explore this we studied the behavior of the MST clusterization
when applied to constrained systems at high energies. We observed that once
we enter the `vapor-branch', the spectra remain unchanged. In Fig.\ref
{figMstGas}) we show the obtained MST cluster distribution for four
different densities at energies that situate the systems in the vapor-branch
in the respective CC. It can be seen that the MST-spectra go from a U-shape
up to an exponential decaying behavior as the system density decreases.
Notice that for densities around $\rho \sim \rho _{fo}^{-}$ a power-law like
shape can be recognized. That means that from the spatial point of view the
system admits, at least in principle, fragment configurations formed by well
separated clusters of almost all possible sizes.

In Fig.\ref{figMstOrtiz}) we show the mean value of the size of the maximum
MST-fragment as a function of the energy for the two densities taken as a
reference. It can be seen that for the dense case, $\rho =\rho _{fo}^{+}$,
the maximum fragment comprises almost all of the mass of the system
irrespective of the value of the energy deposited in it. On the other hand
for $\rho =\rho _{fo}^{-}$ the size of the biggest fragment decreases with
increasing energy. This results support the idea that for the high density
constrained case, no surfaces can be built in order to get small
MST-fragments at high densities, whereas at low densities MST-fragments are
well defined structures in configuration space and the surfaces appear.

We will now discuss the results obtained within the ECFM model. In Fig.\ref
{figecramst}) we show the ECRA and MST mass distribution for the
unconstrained system ,both calculated at $\tau _{ff}$. We also show the same
quantities calculated for the constrained case at densities $\rho =\rho
_{fo}^{+}$ and $\rho =\rho _{fo}^{-}$.

First we can notice that for the dense constrained case MST always gives
essentially the same U shaped curve, while the ECRA spectra show the usual
transition from U shaped to exponentially decaying behavior. On the other
hand, for the low density constrained case and the unconstrained case both
the MST and ECRA results show such a transition.

It is interesting to note that when comparing the most energetic case the
ECRA mass distributions corresponding to the constrained cases present a
steeper slope than the corresponding to unconstrained situation (figures~\ref
{figecramst} (d), (h), and (l)). This happens as a direct consequence of the
existence of boundaries. For the constrained cases the violent inter
particle collisions, that are present at all times due to the reflections of
particles with the constraining walls, destroy almost every correlation in
phase space. On the other hand, for the unconstrained case the system
develops a collective expansive motion which limits the `thermal' component
of the kinetic energy, allowing for the build up of well correlated density
fluctuations in phase space, i.e.ECRA-clusters (note also that
while the ECRA clusters are microscopically stable
for the unconstrained case, the ones for the constrained system are not). 
It is in this regime where the differences between a local equilibrium picture
and a global one become important.

As long as the collective expansive energy in the free case can be neglected
($E\lesssim 0.5\epsilon $) the ECRA spectra present a rather similar shape
for the three shown cases. Moreover, a clear signature of a first order
phase transition can be recognized in the shape of the ECRA spectra noticing
the simultaneous presence of a `liquid-phase' (mass fragments $m\sim 30$)
and a `gas-phase' (mass fragments $m<10$) for $0.1\epsilon \leq E\leq
0.4\epsilon $

This findings clarify the shapes of the corresponding CC. For the dense
constrained system, $\rho _{fo}^{+}$, we have a big configurational
MST-cluster all the time, inside of which an (unstable) ECRA partition can
be defined  that exhibits features expected for a first order phase
transition. The fact that the whole system is strongly interacting (there
are no spatial surfaces dividing ECRA-phases) is responsible of the
reduction of the back-bending behavior in the corresponding CC and the lack
of strong kinetic energy fluctuations. For the more dilute case, $\rho
_{fo}^{-}$, the system is capable of `reproduce' phase space surfaces (at
least at some degree) in configurational space and the mentioned signals
start to be noticeable.

An extra support to the idea that fragmentation do take place in phase space
not only for the free expanding system (see \cite{ale_time}) but also for
the constrained case can be gained if we remove the external `wall'
potential and let the system expand. In Fig.\ref{figFreeVol}) we compare the
asymptotic spectra resulting from such a process of removal of the
constraints, with the ECRA one obtained with boundaries for $\rho _{fo}^{-}$
and $\rho _{fo}^{+}$ cases at $E=0.4\epsilon $ and $E=0.7\epsilon $ . A
quite remarkable agreement can be found. The fact that this agreement is
better for lower densities reflects that different degree of
spatial-fragmentation is achieved at different densities. In this way, for $%
\rho _{fo}^{+}$, the differences in both spectra (slight suppression of
intermediate mass fragments favoring bigger clusters) can be understood as
consequence of an aggregation-like process occurring during the early
expansive stage due to the strong interacting nature of the system.

In Section~\ref{secTRF} we showed for the confined system that the diverging
behavior of the TRF is related to abnormally large fluctuations of the
kinetic energy. In order to understand the origin of those fluctuations we
show in Fig.\ref{figCorrel}) the system kinetic energy $K$ and the mass of
the biggest ECRA-cluster as a function of time for $\rho =\rho _{fo}^{-}$ at 
$E=0.4\epsilon $. Also the mass of the biggest MST-cluster is shown. It is
clearly seen that the first two quantities are well correlated while the MST
biggest cluster is not. This indicates that from a microscopic point of view
the systems is sometimes mainly liquid (big ECRA biggest cluster) and others
it is mainly vapor (small ECRA biggest cluster). This phase alternation is
the kind of coexistence expected for finite system (see~\cite{labestie}).

\section{Conclusions}

\label{secConclu}

In this work we have presented a complete analysis of Caloric Curves,
Thermal Response Functions and MST and ECRA fragment distributions for
constrained and unconstrained excited liquid drops. We have shown that the
effect of constraints is quite important, and that its main result is to
allow the system to reach thermal equilibrium. In particular the Caloric
Curves for constrained cases show a vapor branch which is clearly an
`equilibrium effect' that is not present for unconstrained systems.

We have also verified, studying the behavior of ECRA-cluster mass
distributions,  that fragmentation does occur in phase space in all the
studied cases.  Moreover, phase coexistence can be found in all ECRA spectra
at appropriate energies (we have shown that the ECRA fragments in
constrained systems correspond closely to the MST asymptotic clusters when
the constraints are removed. This reinforces the validity of the ECFM-ECRA
approach to study fragmentation phenomena occurring in finite volumes).

It is interesting to notice that for the constrained case, as the system
density is increased, the expected signals of a first order phase transition
in the CC, TRF and kinetic energy fluctuations are gradually smoothed, and
eventually disappear. This can be associated to the fact that at low
densities ''internal surfaces'' can be developed in the constrained system
allowing the transition to be traced in configurational space (i.e. MST
clusters are formed reflecting the fact that well separated  aggregates
appear in coordinate space).  Such a feature is not possible in the dense
case, all of the time the system is composed by a big configurational
cluster that comprises more than 95\% of the total mass.

Moreover for the constrained case, a relation between the fluctuations in
size of the maximum ECRA-fragment and the kinetic energy fluctuations was
established. As a consequence, the presence of abnormally large fluctuations
in the system kinetic energy could directly be linked with a phase
coexistence phenomenon taking place in a finite system.

As a final remark we want to emphasize that, even some similarities can be
found between the behavior of low density constrained systems and
unconstrained ones at low energies, such a comparison fails at and above the
energy that corresponds to the onset of the fragmentation process, i.e. when
the collective radial modes begin to drive the evolution of the system.

As we mentioned, the role of this collective motion is to behave as a heat
sink, precluding the system from developing a vapor branch, and freezing the
most bound density fluctuations in phase space allowing them to become the
asymptotic clusters. This is why a `local' equilibrium picture, and not a
`global' one, is necessary to describe the fragmentation process correctly,
i.e. taking into account the effects of the radial flux. This teaches us
that when dealing with fragmentation phenomena of the kind appearing in
nuclear multifragmentation, caloric curves which display vapor branches
should be taken with caution, and consequently the results obtained from
models that display such a feature should be at least critically reexamined.

\subsection*{Acknowledgment}

This work was done under partial financial support from the University of
Buenos Aires via Grant No.TW98, and CONICET via Grant No. PIP 4436/96.
A.Chernomoretz acknowledges CONICET for financial support and the warm
hospitality of the Laboratoire de Physique Nucl\'eaire, D\'epartement de 
Physique, Universit\'e Laval (Quebec, Canada).
M.Ison aknowledges finantial support from UBA via a student scholarship. 

\newpage

\subsection{\protect\smallskip Appendix}

Once the clusters have been calculated using the ECFM we have to determine
the time at which the fragmentation is over and the systems enters an
evaporation regime (i.e the fragmentation pattern is formed and the fragments
undergo a simple evaporation process). We will then define the time of
fragment formation to that time at which the MBDF attain microscopic
stability. This property of the clusters has been calculated using what we
call the Short Time Persistence, in which we calculate the stability of MBDF
against evaporation and coalescence. We define the time of stabilization in
the following way: Given a configuration resulting from the ECRA analysis at
a given time $t$ we analyze the microscopic stability of each fragment $%
C_{i}^{t}$ of size $N_{i}^{t}$ by searching on all the fragments $%
C_{j}^{t+dt}$ present at time $t+dt$ for the biggest subset $N_{\max
}]_{i}^{t+dt}$ of particles that belonged to $C_{i}^{t}.$ We then we assign
to this fragment a value $STP_{d}=$ $\frac{N_{\max }]_{i}^{t+dt}}{N_{i}^{t}}%
. $ In this way we are taking into account what we call the ''evaporation
process''. We also have to take into account that there can be some
realization for which cases the $N_{\max }]_{i}^{t+dt}$ does not constitute
a subset of the original cluster $C_{i}^{t}$ but is embedded in a bigger
fragment of mass $N_{i}^{t+dt}$. We include this effect by defining $STP_{i}=%
\frac{N_{\max }]_{i}^{t+dt}}{N_{i}^{t+dt}}$ (with $i$ standing for $inverse)$%
. Finally the Short time persistence reads :

\smallskip 
\begin{equation}
STP(t,dt)=\left\langle \left\langle \left[ \frac{STP_{d}(t,dt)+STP_{i}(t,dt)%
}{2}\right] _{j}\right\rangle _{m}\right\rangle _{e}
\end{equation}

where $\left\langle ...\right\rangle _{m}$ is the mass weighted average over
all the fragments with size $N>3.$ And $\left\langle ...\right\rangle _{e}$
is the average over an ensemble of fragmentation events at a given energy E.

A reference value for $STP(t,dt)$ can be obtained considering that the
fragments undergo only a simple evaporative process. In this way we say that
when $STP$ reaches this reference value the system goes from fragmentation
to evaporation. We call this time $\tau _{ff}$.

\newpage 
\begin{figure}[htbp]
\caption{ In the higher figure: local density at the initialization step
(empty symbols) and local density at fragmentation time (solid symbols) as a
function of the center of mass distance. In the lower figure: local
temperature profile at fragmentation time as a function of the center of
mass distance. Circles denote $E=0.0 \epsilon$, squares denote $E=0.5\epsilon
$, up-triangles denote $E=0.7\epsilon$, diamonds denote $E=0.9\epsilon$, and
left-triangles denote $E=1.2\epsilon$. }
\label{figtloc}
\end{figure}

\begin{figure}[htbp]
\caption{ In figure (a): Caloric curve calculated for the expanding system
(solid circles). The empty squares correspond to an estimation of a
`fake-temperature' that does not take into account in a proper way the
collective motion and is calculated simply as a fraction of the total
kinetic energy (see text for details). Figure (b) shows the associated
thermal response function. }
\label{figCC}
\end{figure}

\begin{figure}[htbp]
\caption{ Temperature of the constrained system as a function of its energy
and density. }
\label{figCCV3D}
\end{figure}

\begin{figure}[htbp]
\caption{ Caloric curve for the constrained system, figure (a), and the
respective thermal response functions, figure (b), for three different
densities: $\rho=\rho_{fo}^{-}$, $\rho_{fo}^{0}$, and $\rho_{fo}^{+}$
(circle, triangle, square symbols in figure (a), and full, dashed,
dotted-dashed lines in figure (b), respectively) }
\label{figCCV}
\end{figure}


\begin{figure}[htbp]
\caption{ Relative kinetic fluctuation, $A(E)$, for the constrained case
calculated for different densities, as a function of the total energy. The
symbols correspond to direct calculations for $\rho=\rho^+_{fo}$ (empty
squares) and for $\rho=\rho^-_{fo}$ (solid circles). The lines correspond to 
$A$ estimations using the caloric curve (solid-line for $\rho=\rho^-_{fo}$,
and dashed-line for $\rho=\rho^+_{fo}$). The dashed-dotted line shows the
canonical expectation value for $A$. }
\label{figFlucV}
\end{figure}

\begin{figure}[htbp]
\caption{ Relative kinetic energy fluctuation, for the non constrained
system, as a function of the core-energy. The calculation includes only
events with $N_{core}=60\pm5$ particles inside the core. }
\label{figFluc}
\end{figure}

\begin{figure}[htbp]
\caption{MST mass distribution calculated for the vapor-branch of
constrained systems of densities: $\rho= 0.08\sigma^{-3}(\rho^+_{fo})$, $%
0.035\sigma^{_3}$, $0.02\sigma^{-3} (\rho^-_{fo})$, and $0.01\sigma^{-3}$
are shown in figures (a), (b), (c), and (d) respectively. }
\label{figMstGas}
\end{figure}

\begin{figure}[htbp]
\caption{ Mean value of the mass of the biggest MST-cluster for the
constrained case. Circles denote $\rho=\rho_{fo}^+$, and triangles $%
\rho=\rho_{fo}^-$ }
\label{figMstOrtiz}
\end{figure}

\begin{figure}[htbp]
\caption{ ECRA (solid circles) and MST (empty squares) mass spectra
calculated at fragmentation time $\tau_{ff}$. The first column corresponds
to the free expanding case, the others to the $\rho=\rho^-_{fo}$ and $%
\rho=\rho^+_{fo}$ confined cases respectively. The considered system
energies were: $E=-0.5\epsilon$, $E=0.0\epsilon$, $E=0.5\epsilon$, $%
E=1.2\epsilon$, in figures (a)-(d) for the free expanding case, and $%
E=-0.2\epsilon$, $E=0.1\epsilon$, $E=0.4\epsilon$, $E=1.2\epsilon$ in
figures (e)-(f) and (i)-(l) for the constrained cases.}
\label{figecramst}
\end{figure}

\begin{figure}[htbp]
\caption{ Asymptotic mass distribution after the `wall' removal (empty
squares) and ECRA spectra calculated within boundaries (solid circles) for
densities $\rho=\rho^+_{fo}$ (first raw), and $\rho^-_{fo}$ (second raw).
Figures (a) and (c) corresponds to $E=0.4\epsilon$, while (b) and (d) to $%
E=0.7\epsilon$ }
\label{figFreeVol}
\end{figure}

\begin{figure}[htbp]
\caption{In figure (a), the system kinetic energy is shown as a function of
time for the constrained case at $\rho=\rho^-_{fo}$ and $E=0.4\epsilon$. The
temporal dependence of the mass of the biggest ECRA and MST clusters are
shown in figures (b) and (c) respectively.}
\label{figCorrel}
\end{figure}


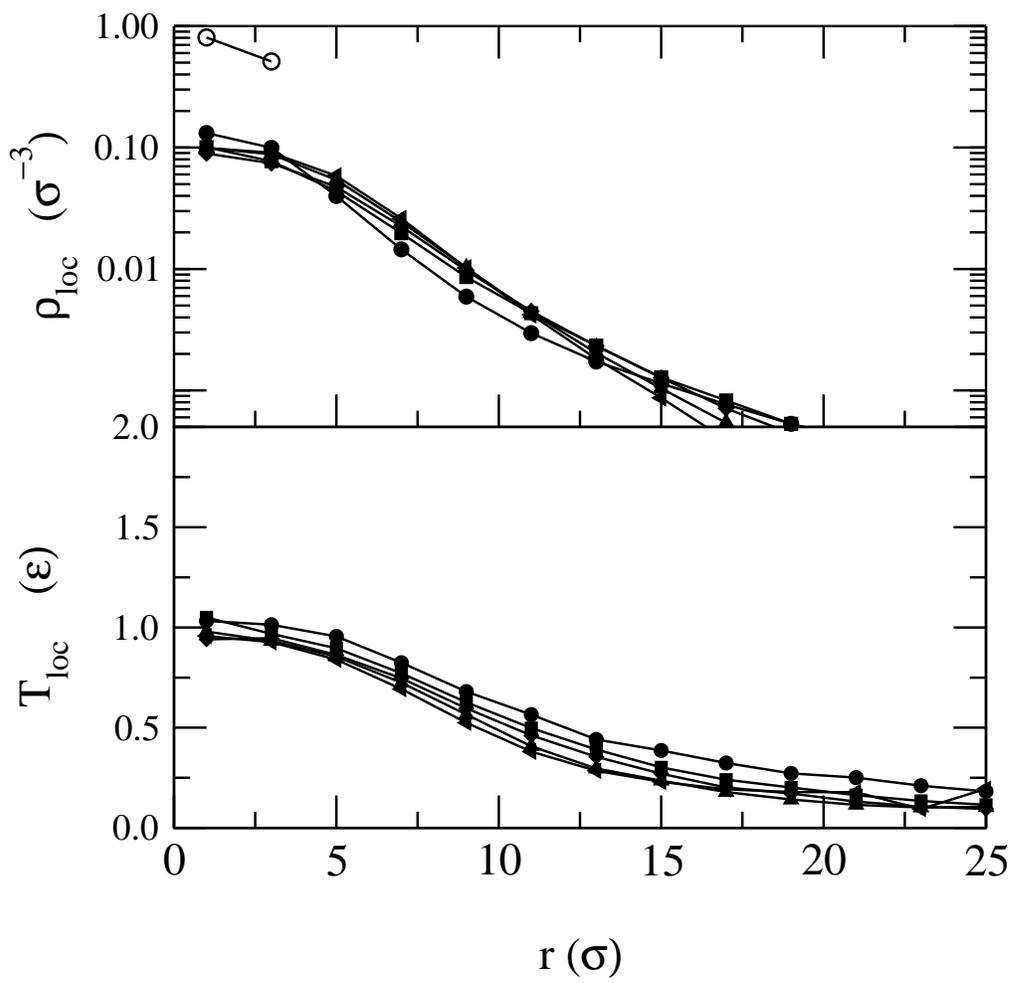
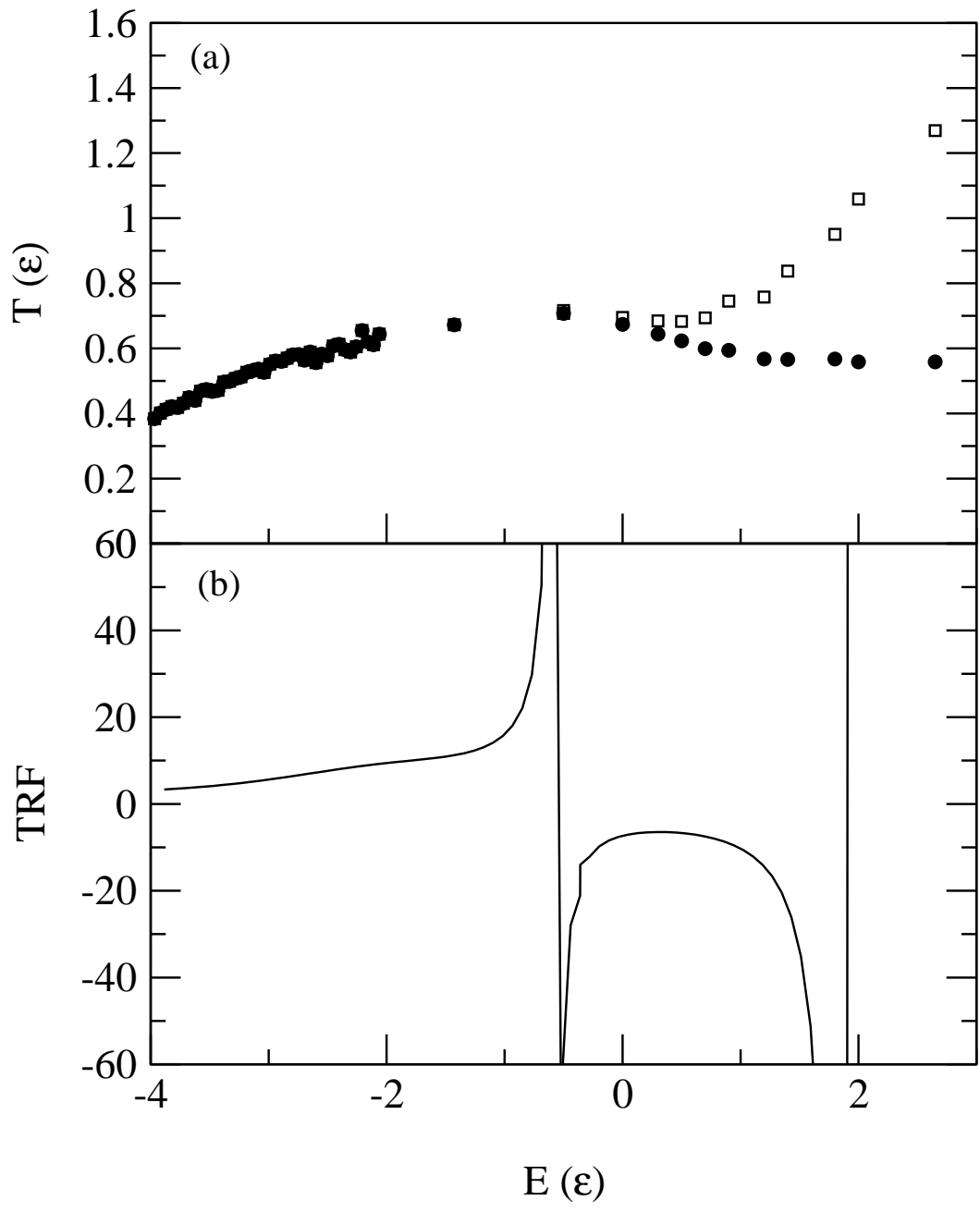
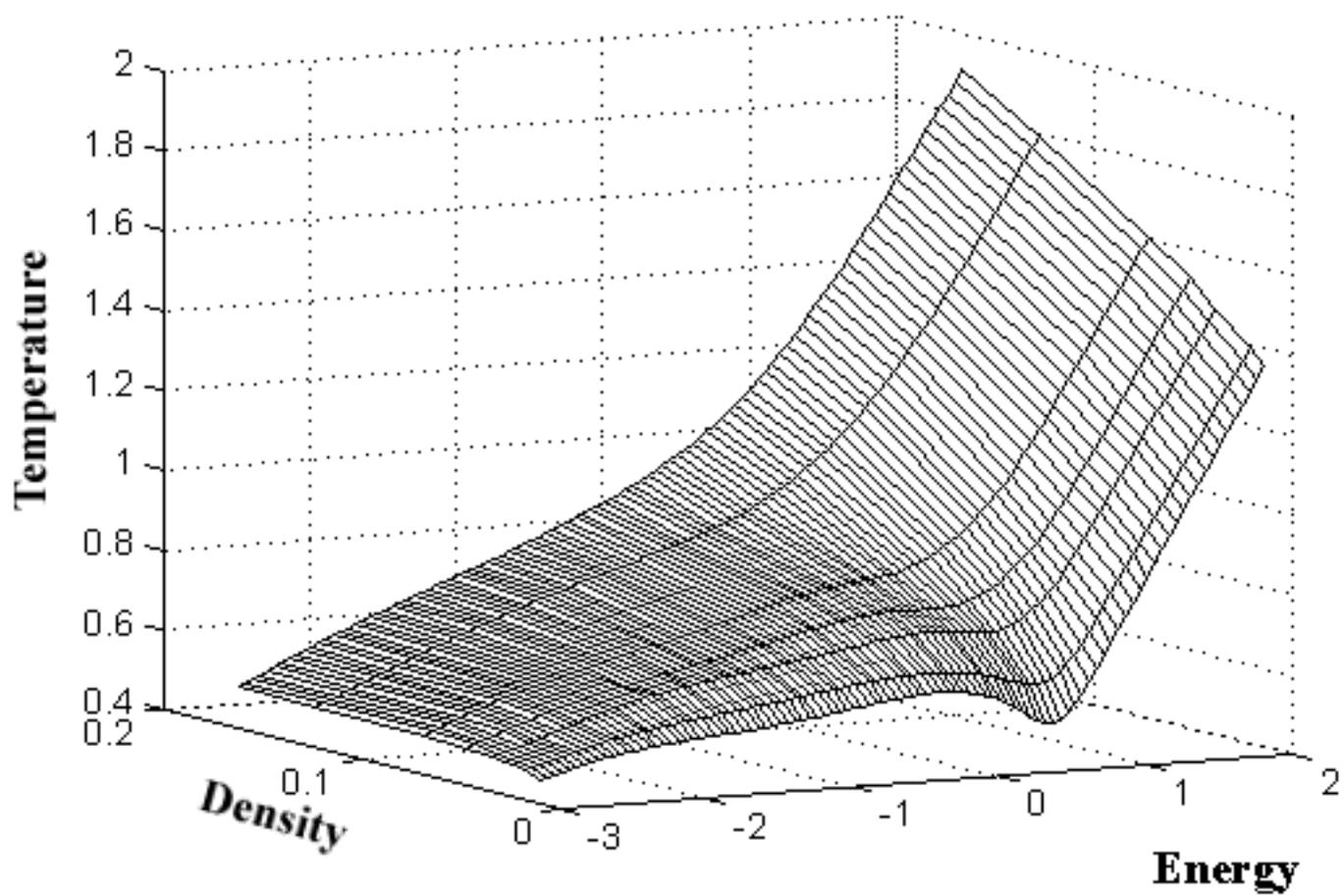
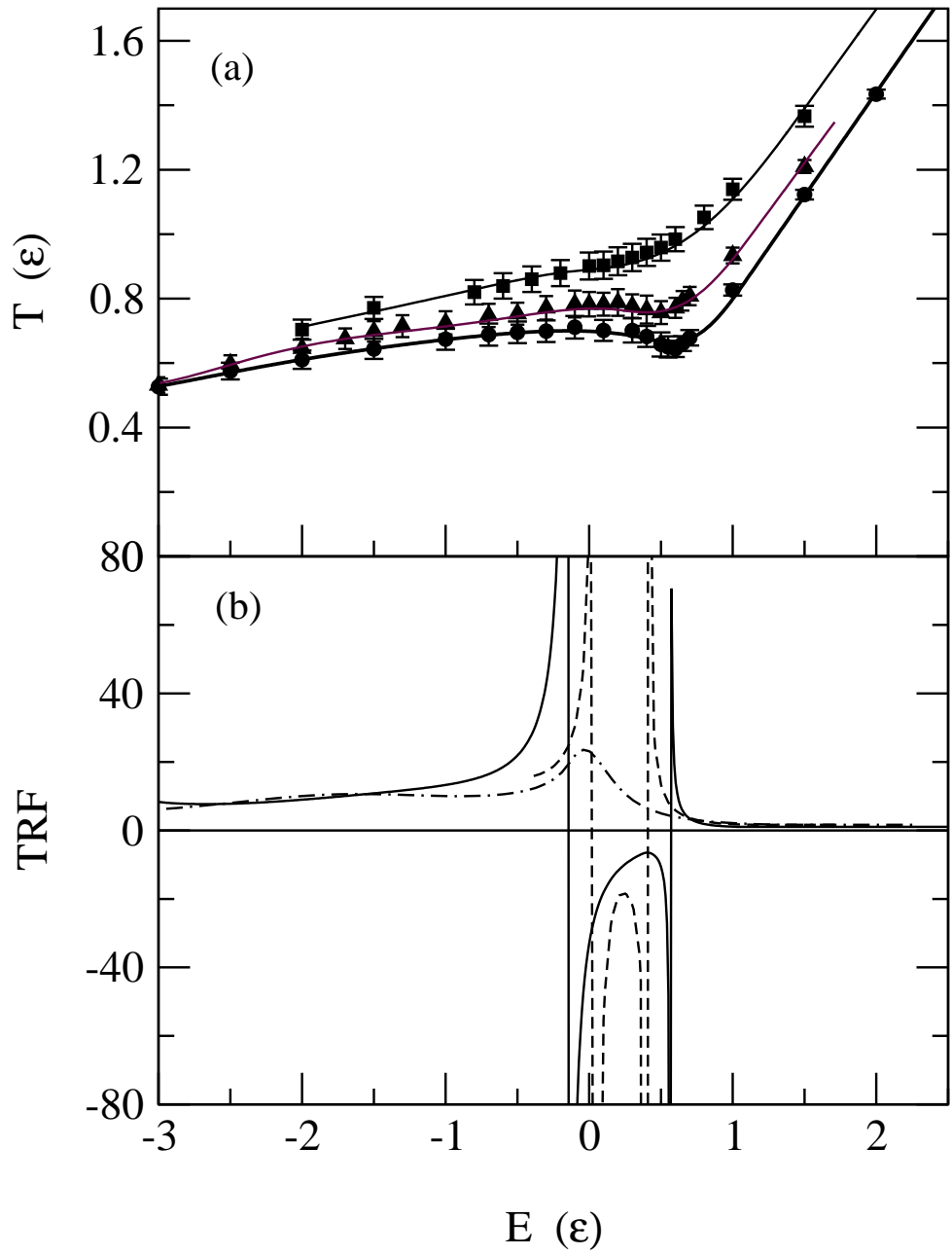
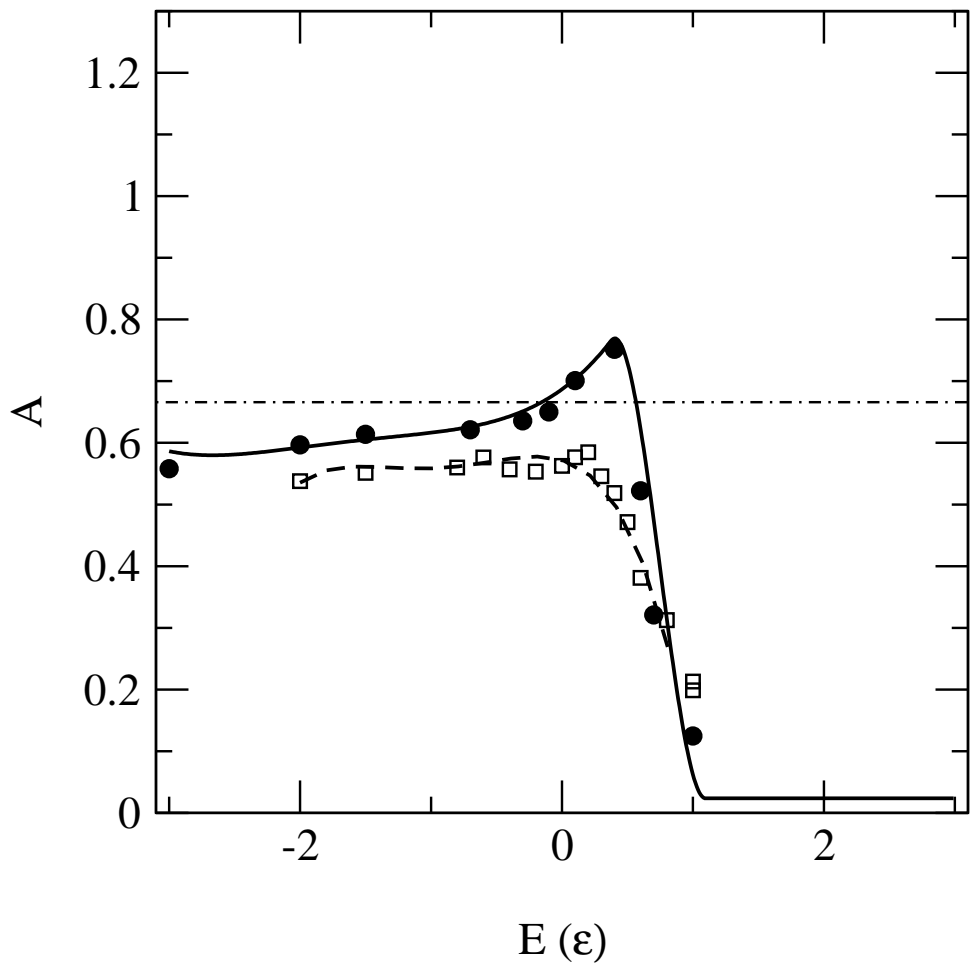
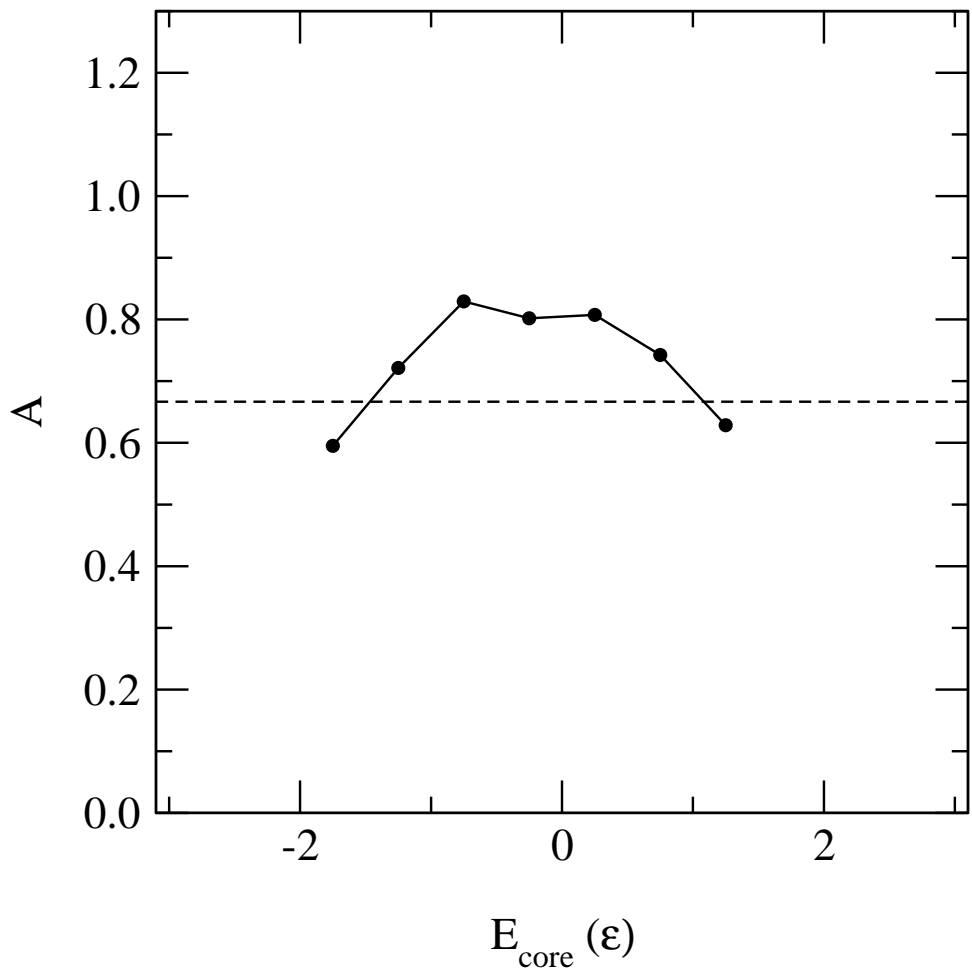
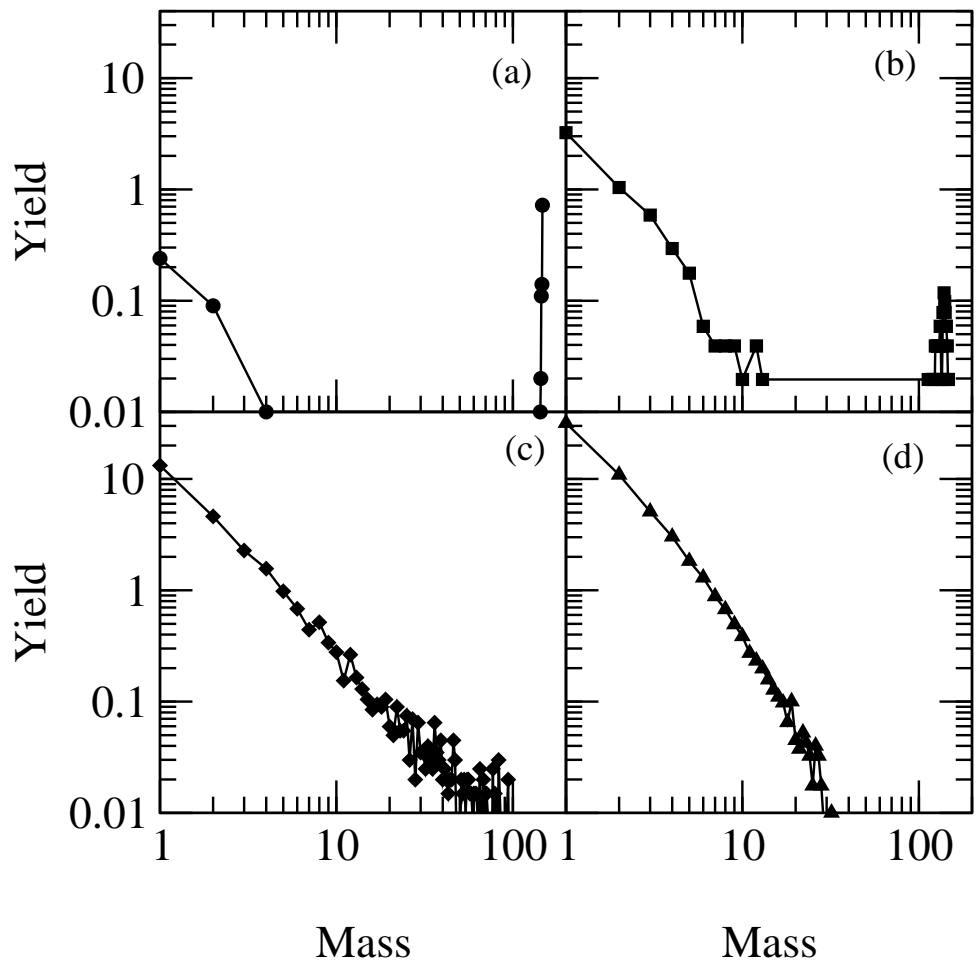
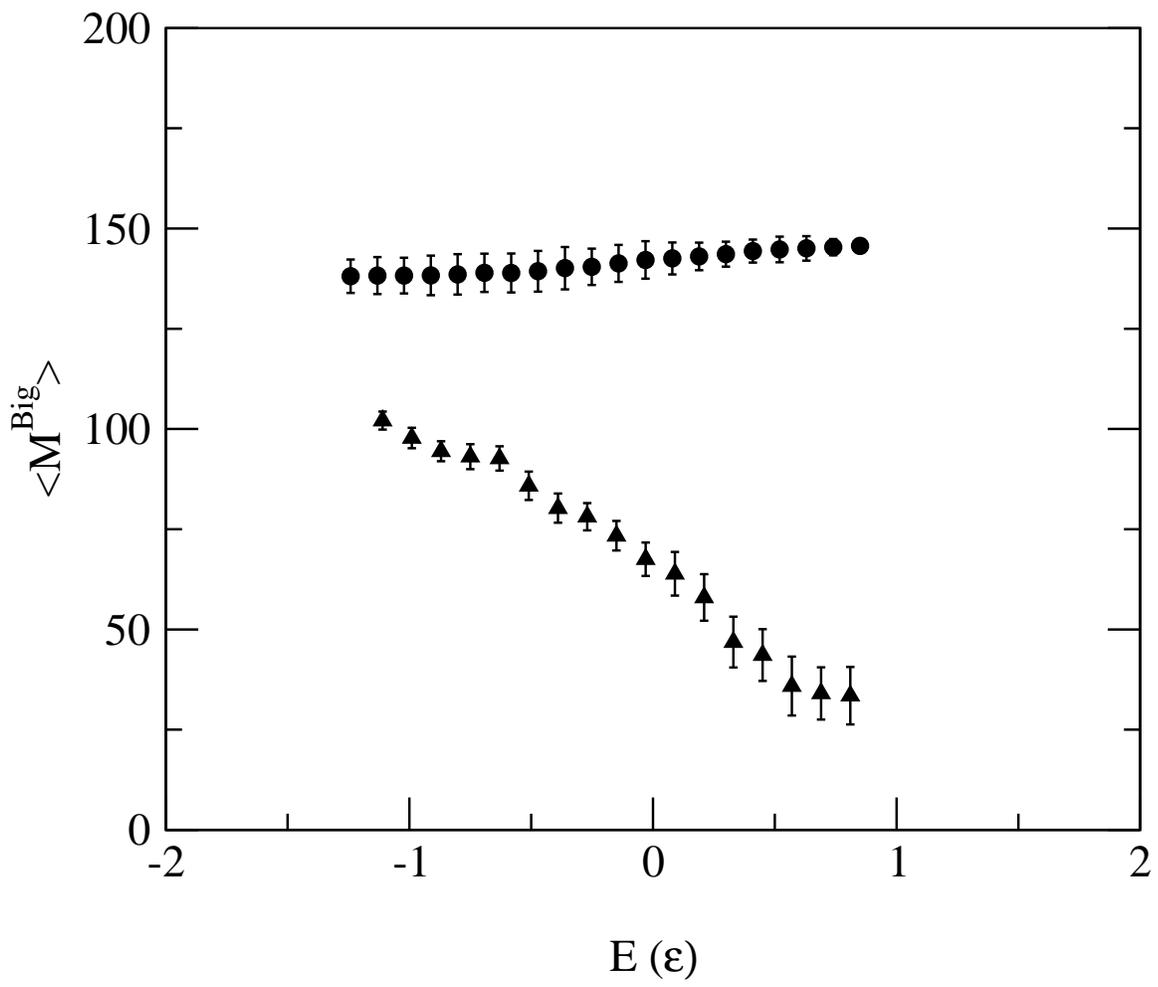
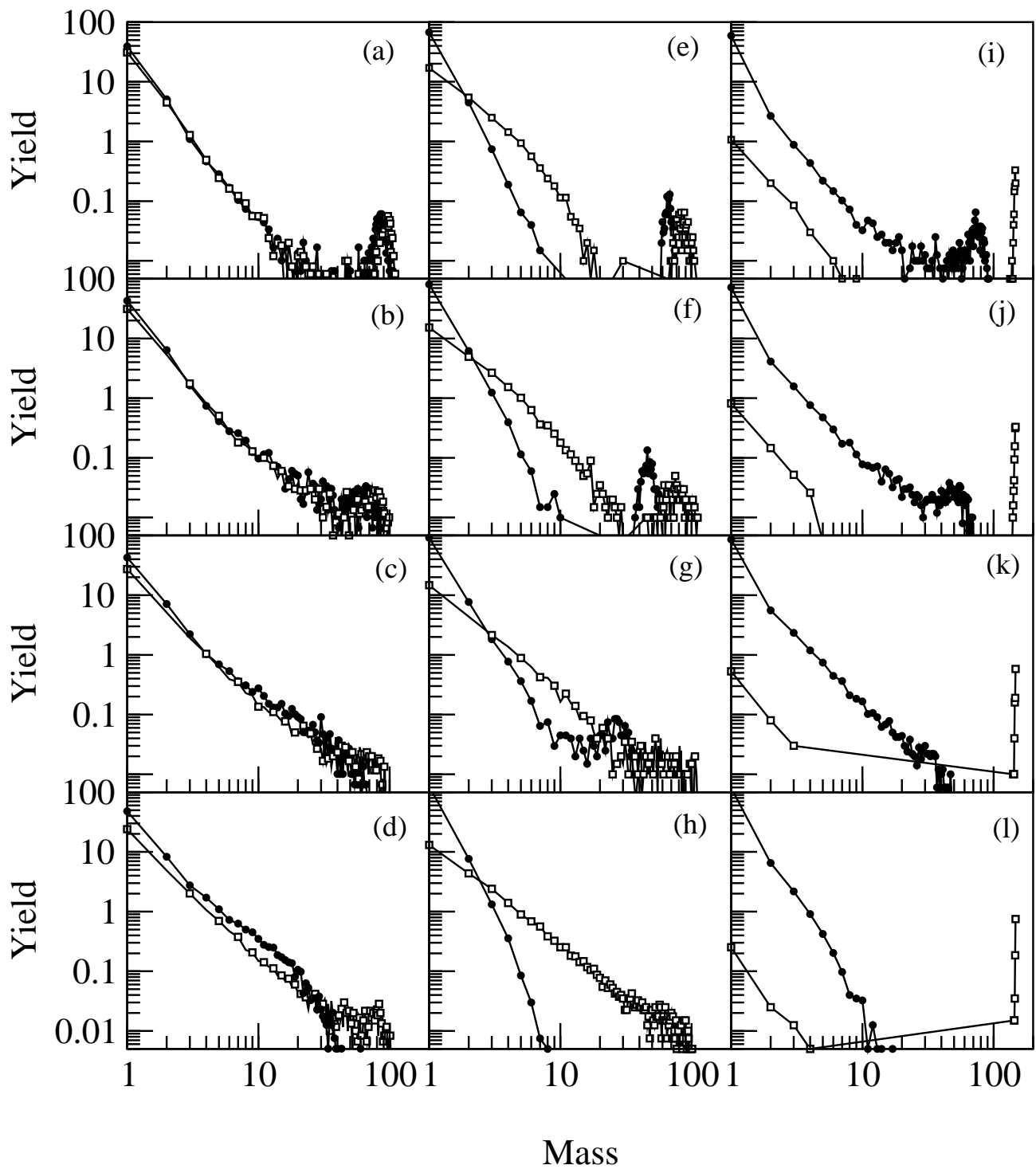
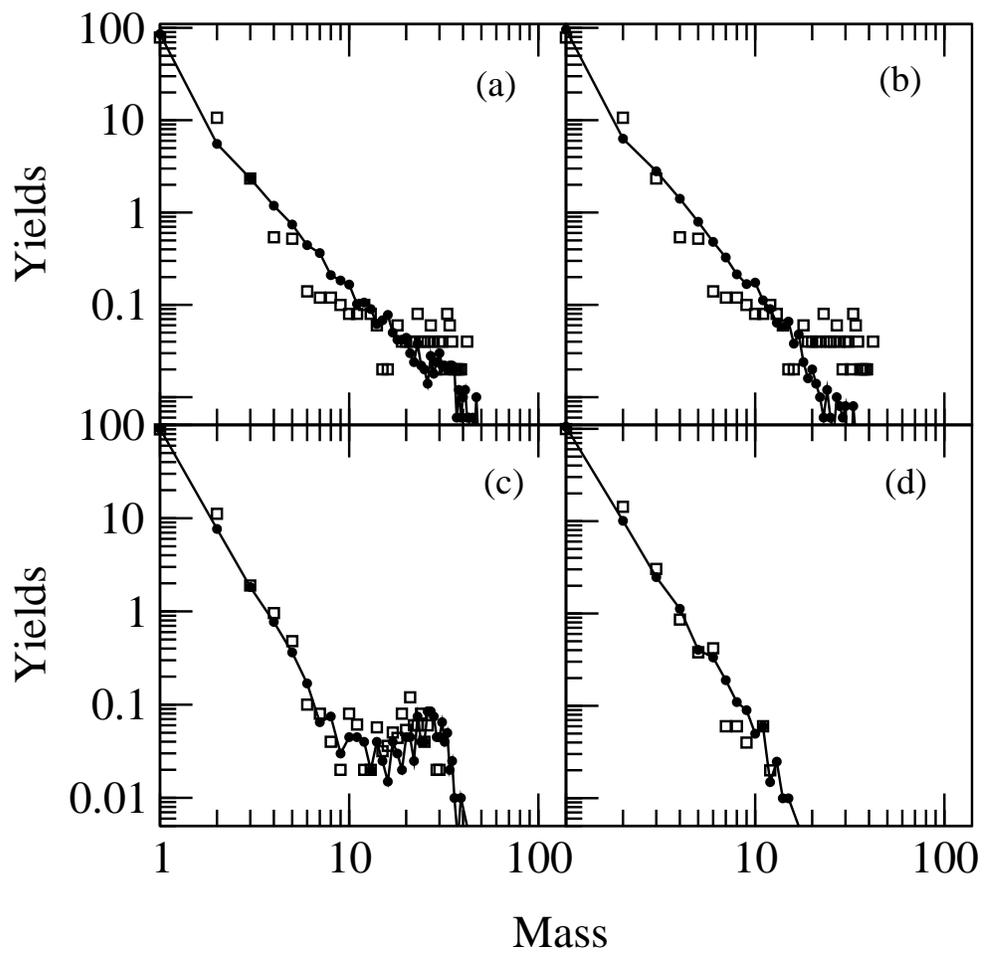
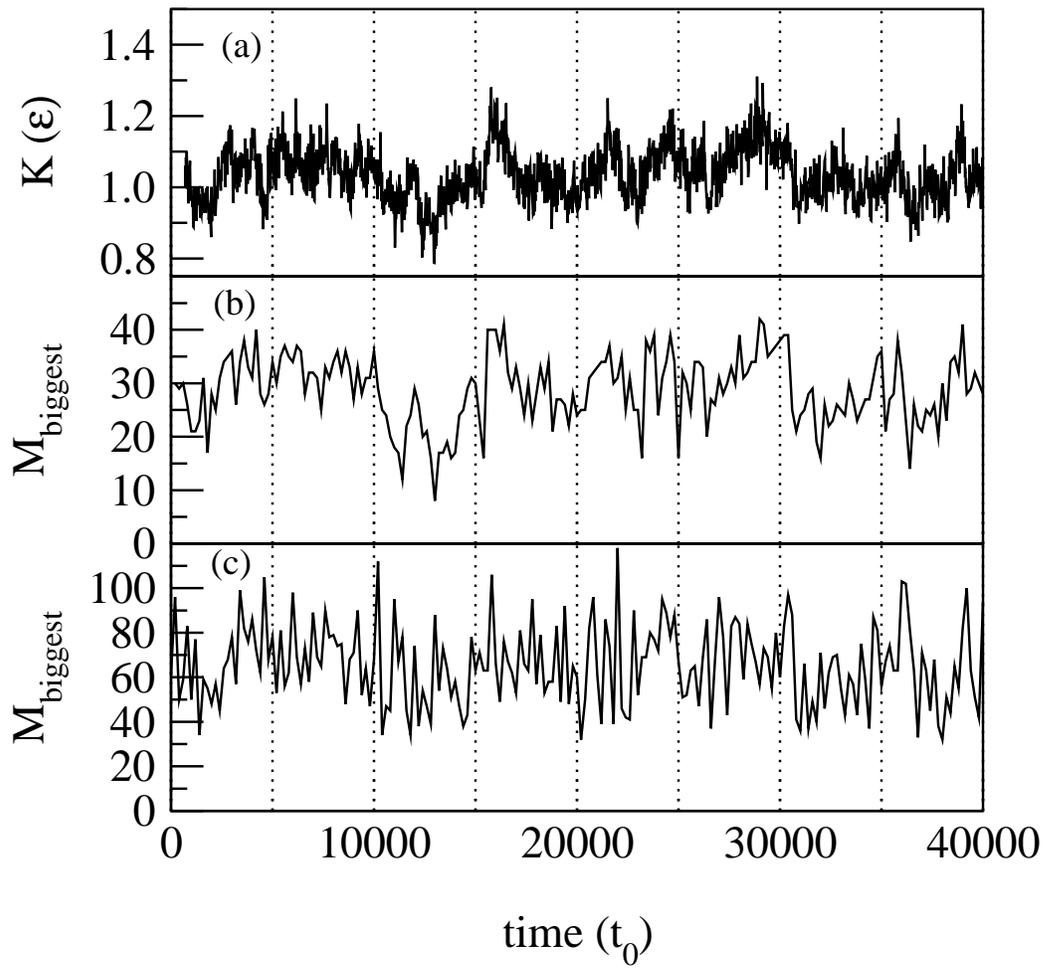
\end{document}